\documentclass[conference]{IEEEtran}
\IEEEoverridecommandlockouts
\usepackage{cite}
\usepackage{amsmath,amssymb,amsfonts}
\usepackage{algorithmic}
\usepackage{graphicx}
\usepackage{textcomp}
\usepackage{xcolor}
\usepackage{booktabs}
\usepackage{multirow}
\def\BibTeX{{\rm B\kern-.05em{\sc i\kern-.025em b}\kern-.08em
    T\kern-.1667em\lower.7ex\hbox{E}\kern-.125emX}}

\usepackage{marvosym}  

\begin{document}

\title{Spatial-Aware Conditioned Fusion for Audio-Visual Navigation}

\author{
Shaohang Wu$^{1,2,3}$, and Yinfeng Yu$^{1,2,3}$$^{,\mbox{\Letter}}$%
\thanks{$^{\mbox{\Letter}}$Yinfeng Yu is the corresponding author(E-mail: yuyinfeng@xju.edu.cn).}
\\
$^1$Joint Research Laboratory for Embodied Intelligence, Xinjiang University\\
$^2$Joint International Research Laboratory of Silk Road Multilingual Cognitive Computing, Xinjiang University\\
$^3$School of Computer Science and Technology, Xinjiang University, Urumqi 830017, China%
}

\maketitle

\begin{abstract}

Audio-visual navigation tasks require agents to locate and navigate toward continuously vocalizing targets using only visual observations and acoustic cues. However, existing methods mainly rely on simple feature concatenation or late fusion, and lack an explicit discrete representation of the target’s relative position, which limits learning efficiency and generalization. We propose Spatial-Aware Conditioned Fusion (SACF). SACF first discretizes the target’s relative direction and distance from audio-visual cues, predicts their distributions, and encodes them as a compact descriptor for policy conditioning and state modeling. Then, SACF uses audio embeddings and spatial descriptors to generate channel-wise scaling and bias to modulate visual features via conditional linear transformation, producing target-oriented fused representations.  SACF improves navigation efficiency with lower computational overhead and generalizes well to unheard target sounds.

\end{abstract}


\section{Introduction}
\begin{figure}[htbp]
\includegraphics[width=1.1\columnwidth]{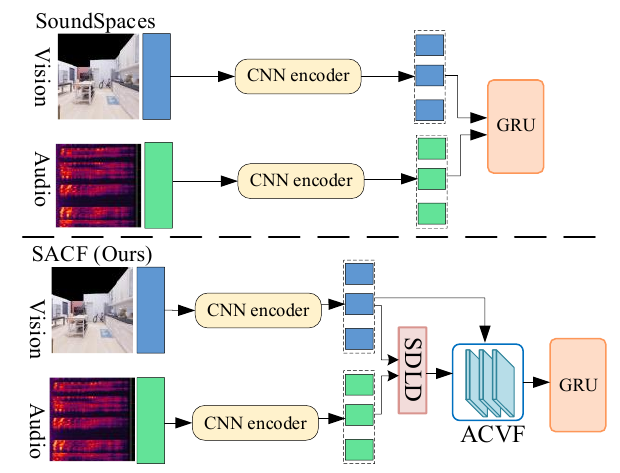} 
\caption{Comparison between the SoundSpaces baseline and our method.}
\label{fig1}
\end{figure}

Audio-visual navigation (AVN) utilizes audio-visual cues to localize sound sources \cite{b2,b3}. Existing methods predominantly employ simple feature concatenation \cite{b2,b4} or late fusion, as illustrated in Fig.~\ref{fig1}. However, such fusion approaches often overlook two core issues: First, they lack explicit spatial representation. Many methods treat audio more as a directional cue or implicit hint, requiring the policy network to learn the target's location autonomously within a high-dimensional continuous feature space \cite{yu2025dynamic}. Second, visual representation learning lacks auditory-conditioned guidance. Visual encoders typically treat the entire field of view uniformly, struggling to dynamically emphasize target-relevant visual information based on sound source cues \cite{zhang2025advancing,li2025audio}. These shortcomings limit model learning efficiency and result in poor generalization when testing in unfamiliar environments or without sound.

Human navigation in the real world first relies on binaural auditory differences to roughly determine the direction and distance of sound sources. This spatial intent then guides visual attention to search for paths in specific directions. Inspired by this, this paper proposes a novel framework called Spatial-Aware Conditioned Fusion (SACF). It aims to address the aforementioned challenges through explicit spatial discretization representation and deep cross-modal modulation. SACF comprises two core components: 1) Spatially Discretized Localization Descriptor (SDLD). Unlike traditional regression-based predictions, this module discretizes the relative position of sound sources into intervals of direction and distance, predicts their probability distributions, and encodes them into compact geometric descriptors. This approach not only reduces learning complexity but also provides robust spatial priors for subsequent decision-making. 2) To achieve deep auditory guidance for visual processing, we propose Audio-Descriptor Conditioned Visual Fusion (ACVF). This mechanism avoids simple feature concatenation by dynamically generating per-channel scaling and bias parameters, conditioned on audio embeddings and the generated SDLD. These parameters transform visual feature maps through linear modulation, thereby suppressing irrelevant background noise and enhancing target-oriented visual representations during feature extraction.

Experimental results demonstrate that SACF not only enhances overall navigation performance but also significantly improves learning efficiency and cross-scenario generalization capabilities. Specifically, ACVF employs FiLM-style channel modulation \cite{b9} instead of spatial weighting, achieving efficient cross-modal fusion with lower computational and parameter costs. Ablation studies further validate the contributions of SDLD spatial priors and ACVF modulation mechanisms to performance gains.

The main contributions of this paper are summarized as follows:
\begin{itemize}
\item We designed the SDLD module to model the relative positions of sound sources in a direction-distance distribution format and encode them into compact descriptors for policy conditioning and state modeling.

\item We proposed the ACVF mechanism, which dynamically modulates visual features based on spatial intent, significantly enhancing the directionality and effectiveness of visual representations in navigation tasks.

\item We evaluated this approach on SoundSpaces, achieving favourable navigation efficiency on both the Replica and Matterport3D datasets, it exhibited  exceptional generalization capabilities particularly in cross-scene tests where the target sound is unheard.
\end{itemize}

\section{Related Work}

\subsection{Audio-Visual Navigation Integration Mechanism}

Audio-visual navigation (AVN) is gaining increasing attention as a key task in embodied intelligence \cite{yu2025dope}. SoundSpaces \cite{b2} provides a standardized simulation platform and benchmark. Early approaches typically concatenate audio and visual features and feed them into a recursive policy \cite{b2}, but this leads to limited cross-modal interactions. Subsequent work has advanced AVN primarily through two directions: fusion and memory modeling. Regarding fusion, attention-based approaches like FSAAVN \cite{yu2022pay} dynamically adjust modal weights \cite{yu2025dgfnet}. For memory, structured acoustic memory or explicit history aggregation helps preserve sound source cues over longer time scales \cite{b11}. Additionally, some approaches incorporate active acoustic emission and echo analysis to handle sparse or missing sources \cite{yu2023echo,yusound}, while others leverage multi-agent collaborative acoustic measurements to enhance estimation accuracy \cite{yu2023measuring}. However, existing methods still lack explicit spatial structure representation and auditory-guided visual learning. To address these limitations, our SACF significantly enhances learning efficiency and generalization capabilities through explicit spatial discretization and channel-level conditional modulation.
\subsection{Channel-Wise Conditional Modulation}
FiLM (Feature-Level Linear Modulation) was initially developed for conditional representation learning in visual reasoning tasks. This model generates channel-level scaling and bias parameters $(\gamma,\beta)$ from conditional inputs, then modulates intermediate features via $\gamma \odot x + \beta$ to achieve lightweight conditional feature selection\cite{b9}. We observe that this channel-level modulation effectively adjusts feature weights with low overhead and good trainability. In contrast, attention-based cross-modal fusion typically relies on spatial weights or token-level interactions \cite{yu2022pay,b14,yu2021weavenet,zhang2025iterative}, which increase computational/parameter overhead and may complicate optimization in reinforcement learning. FiLM-style modulation avoids explicit spatial reweighting, instead performing conditional selection at the channel dimension. This makes it more concise and efficient, better suited for fusion tasks in reinforcement learning \cite{b9,b15}.

\begin{figure*}[htb!] 
\centering
\includegraphics[width=\textwidth]{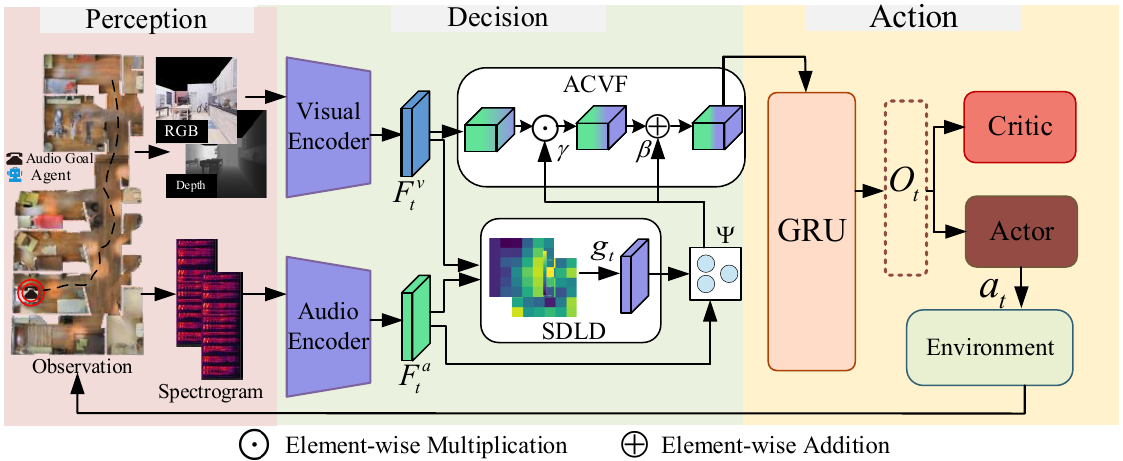} 
\caption{SACF for audio-visual navigation receives image and audio observations $o_t=\{I_t, D_t, A_t\}$ at each time step. Features are extracted via visual/audio encoders, and spatial descriptors $g_t$ are generated by SDLD. Subsequently, ACVF conditions on $[F_t^a; g_t]$ to generate FiLM-style channel modulation parameters that enhance the directionality of visual representations. These fused features are input into a GRU, which outputs action $a_t$ and value estimates via Actor-Critic for closed-loop decision-making.}
\label{fig2}
\end{figure*}

\section{Methodology}
As shown in Fig.~\ref{fig2}, this paper proposes the Spatial-Aware Conditioned Fusion (SACF) framework for audio-visual navigation tasks. The overall process consists of three stages: perception, decision, and action. During the perception stage, the agent interacts with the environment to obtain first-person multimodal observations $o_t=\{I_t, D_t, A_t\}$ at each time step $t$, where $I_t$ and $D_t$ represent RGB and depth images, respectively, and $A_t$ denotes the audio observation. During the decision phase, the visual and auditory information obtained from observations are processed through visual and auditory encoders, respectively, yielding visual feature maps $F_t^v$ and audio feature maps $F_t^a$. Subsequently, Spatially Discretized Localization Descriptor (SDLD) module processes these to yield the spatial descriptor vector $g_t$. Building upon this, the Audio-Descriptor Conditioned Visual Fusion (ACVF) module generates FiLM-style channel modulation parameters conditioned on $\left[F_t^a; g_t\right]$, producing a more directional fused visual representation. Finally, the fused representation feeds into a Gated Recurrent Unit (GRU) \cite{b16} to model temporal state, and through Actor-Critic \cite{b17,b18} outputs action ($a_t$) and value estimates. The agent continuously interacts with the environment based on these outputs until the vocalization goal is achieved. The following sections detail the structural design and implementation specifics of SDLD and ACVF respectively.

\subsection{Spatially Discretized Localization Descriptor}\label{AA}
\begin{figure}[htbp]
\includegraphics[width=1.1\columnwidth]{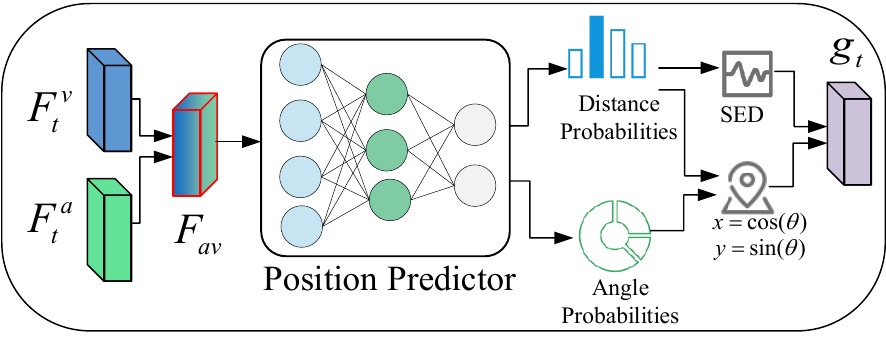} 
\caption{SDLD fuses $F_t^v$ and $F_t^a$ into $F_{av}$ to predict distance/angle distributions, then takes the expectation to obtain $\hat{d},\hat{\theta}$. These are encoded as compact spatial descriptors using $(\cos\hat{\theta},\sin\hat{\theta})$.}
\label{fig3}
\end{figure}

Spatially Discretized Localization Descriptor (SDLD) explicitly infers the relative spatial intent of sound sources from joint visual and auditory cues. It discretizes the target's relative spatial position into direction and distance, predicts its distribution, and encodes it into a compact descriptor for use in policy conditioning and state modeling. First, to obtain a joint representation for localization, we fuse the visual feature map $F_t^v$ with the audio feature map $F_t^a$ to produce the audio-visual joint feature $F_{av}$. As shown in Fig.~\ref{fig3}.

Direct regression on continuous coordinates often struggles to capture the multimodal spatial distribution caused by echoes in real environments and can easily lead to training oscillations. In contrast, discretized representations can explicitly model localization uncertainty through probability distributions. Based on this, we divide the continuous space into $K=20$ discrete categories, a choice driven by computational efficiency and physical granularity. For a 30-meter range, 20 intervals correspond to an average interval size of 1.5 meters. This is roughly equivalent to the typical step size of an agent or the safe obstacle-avoidance distance in indoor navigation, providing sufficient spatial resolution without introducing excessive gaps in the distribution. Distance classification covers the range $0 \sim 30$ meters, while angle classification spans the full range $-\pi \sim +\pi$. The position prediction head outputs non-normalized probabilities for each category, which are then transformed via the Softmax function into smooth probability distributions $P_d$ and $P_\theta$. To achieve more precise continuous value estimates than simple classification, we do not directly select the category center with the highest probability. Instead, we compute the mathematical expectation of the prediction distribution using a weighted average. Let $c_i^d$ and $c_i^\theta$ denote the center values for the $i$th distance and angle category, respectively. The predicted distance $\hat{d}$ and azimuth $\hat{\theta}$ are then: 
\begin{equation}
\hat{d} = \sum_{i=1}^{K} P_d(i) \cdot c_i^d, \quad \hat{\theta} = \sum_{i=1}^{K} P_\theta(i) \cdot c_i^\theta.
\end{equation}
Subsequently, we transform the polar coordinate predictions into direction vectors in Cartesian coordinates to avoid numerical instability caused by angular periodicity:
\begin{equation}
x_t = \cos(\hat{\theta}), \quad y_t = \sin(\hat{\theta}).
\end{equation}
In addition to position coordinates, the module computes a Sound Event Detection (SED) score $s_t$ to determine if the agent has reached the target (e.g., when $\hat{d}$ falls below a threshold). We feed the triplet $(s_t, x_t, y_t)$ as the current observation input into a Long Short-Term Memory (LSTM) network \cite{b19,b20}. The LSTM retains memory of historical observation sequences, dynamically refining the current localization guess using temporal cues (such as the gradient of sound intensity change after the agent moves one step). It ultimately outputs a spatial descriptor $g_t$ that incorporates spatio-temporal consistency.

\begin{table*}[htbp]
\centering
\caption{Performance comparison with other methods under the Depth setting. SPL, SR, SNA are percentages.}
\label{tab:tab1}
\setlength{\tabcolsep}{5pt}
\renewcommand{\arraystretch}{1.15}
\resizebox{\textwidth}{!}{%
\begin{tabular}{l ccc ccc ccc ccc}
\toprule
\multirow{3}{*}{\textbf{Method}}
& \multicolumn{6}{c}{\textbf{Replica}}
& \multicolumn{6}{c}{\textbf{Matterport3D}} \\
\cmidrule(lr){2-7}\cmidrule(lr){8-13}
& \multicolumn{3}{c}{\textbf{Heard}}
& \multicolumn{3}{c}{\textbf{Unheard}}
& \multicolumn{3}{c}{\textbf{Heard}}
& \multicolumn{3}{c}{\textbf{Unheard}} \\
\cmidrule(lr){2-4}\cmidrule(lr){5-7}\cmidrule(lr){8-10}\cmidrule(lr){11-13}
& \textbf{SPL}$\uparrow$ & \textbf{SR}$\uparrow$ & \textbf{SNA}$\uparrow$
& \textbf{SPL}$\uparrow$ & \textbf{SR}$\uparrow$ & \textbf{SNA}$\uparrow$
& \textbf{SPL}$\uparrow$ & \textbf{SR}$\uparrow$ & \textbf{SNA}$\uparrow$
& \textbf{SPL}$\uparrow$ & \textbf{SR}$\uparrow$ & \textbf{SNA}$\uparrow$ \\
\midrule
Random \cite{b11}               & 4.9  & 18.5 & 1.8  & 4.9  & 18.5 & 1.8  & 2.1  & 9.1  & 0.8  & 2.1  & 9.1  & 0.8 \\
Direction Follower \cite{b11}   & 54.7 & 72.0 & 41.1 & 11.1 & 17.2 & 8.4  & 32.3 & 41.2 & 23.8 & 13.9 & 18.0 & 10.7 \\
Frontier Waypoints \cite{b11}   & 44.0 & 63.9 & 35.2 & 6.5  & 14.8 & 5.1  & 30.6 & 42.8 & 22.2 & 10.9 & 16.4 & 8.1 \\
Supervised Waypoints \cite{b11} & 59.1 & 88.1 & 48.5 & 14.1 & 43.1 & 10.1 & 21.0 & 36.2 & 16.2 & 4.1  & 8.8  & 2.9 \\
Gan et al. \cite{b3}          & 57.6 & 83.1 & 47.9 & 7.5  & 15.7 & 5.7  & 22.8 & 37.9 & 17.1 & 5.0  & 10.2 & 3.6 \\
SoundSpaces \cite{b2}          & 74.4 & 91.4 & 48.1 & 34.7 & 50.9 & 16.7 & 54.3 & 67.7 & 31.3 & 21.9 & 33.5 & 10.4 \\
AGSA  \cite{li2025audio}          & 75.5 & 93.2 & \textbf{52.0} & 36.6 & 48.3 & \textbf{22.4} & 54.1 & 70.0 & 30.0 & 26.2 & 36.5 & 13.1 \\
\textbf{SACF (Ours)}     & \textbf{80.3} & \textbf{96.3} & 51.7 & \textbf{43.9} & \textbf{79.1} & 18.7
                        & \textbf{55.0} & \textbf{69.0} & \textbf{32.5} & \textbf{42.4} & \textbf{58.3} & \textbf{28.0} \\
\bottomrule
\end{tabular}%
}
\end{table*}

\begin{table*}[htbp]
\centering
\caption{Performance comparison with other methods under the RGB setting. SPL, SR, SNA are percentages.}
\label{tab:tab2}
\setlength{\tabcolsep}{6pt}
\renewcommand{\arraystretch}{1.15}
\resizebox{\textwidth}{!}{%
\begin{tabular}{l|ccc|ccc|ccc|ccc}
\toprule
\multirow{3}{*}{\textbf{Method}}
& \multicolumn{6}{c|}{\textbf{Replica}}
& \multicolumn{6}{c}{\textbf{Matterport3D}} \\
\cmidrule(lr){2-7}\cmidrule(lr){8-13}
& \multicolumn{3}{c|}{\textbf{Heard}}
& \multicolumn{3}{c|}{\textbf{Unheard}}
& \multicolumn{3}{c|}{\textbf{Heard}}
& \multicolumn{3}{c}{\textbf{Unheard}} \\
\cmidrule(lr){2-4}\cmidrule(lr){5-7}\cmidrule(lr){8-10}\cmidrule(lr){11-13}
& \textbf{SPL}$\uparrow$ & \textbf{SR}$\uparrow$ & \textbf{SNA}$\uparrow$
& \textbf{SPL}$\uparrow$ & \textbf{SR}$\uparrow$ & \textbf{SNA}$\uparrow$
& \textbf{SPL}$\uparrow$ & \textbf{SR}$\uparrow$ & \textbf{SNA}$\uparrow$
& \textbf{SPL}$\uparrow$ & \textbf{SR}$\uparrow$ & \textbf{SNA}$\uparrow$ \\
\midrule
SoundSpaces \cite{b2}
& 62.6 & 72.1 & 31.5
& 24.9 & 35.6 & 14.1
& 44.7 & 64.3 & 22.0
& 20.4 & 30.4 & 7.7 \\
\textbf{SACF (Ours)}
& \textbf{73.5} & \textbf{90.1} & \textbf{38.3}
& \textbf{36.0} & \textbf{62.2} & \textbf{14.2}
& \textbf{45.6} & \textbf{67.7} & \textbf{23.2}
& \textbf{39.6} & \textbf{57.7} & \textbf{22.8} \\
\bottomrule
\end{tabular}%
}
\end{table*}

\subsection{Audio-Descriptor Conditioned Visual Fusion }
To achieve deep auditory guidance of visual perception, we propose the ACVF module. Rather than simply concatenating features, this module dynamically filters visual information using the Feature Linear Modulation (FiLM) [9] mechanism. It conditions on the spatial descriptor $g_t$ generated by SDLD and the global audio feature $F_t^a$.
Unlike spatial attention mechanisms focused on pixel-level alignment, we employ FiLM [9] for channel-level feature modulation. This mechanism dynamically adjusts the weights of visual channels through linear transformations, prioritizing the enhancement of sensitivity to key feature types such as geometric structures and passability based on auditory cues, rather than being confined to specific pixel locations.
This macro-level guidance aligns closely with the navigation task's decision logic of first determining direction followed by fine-grained approach. It enables agents to rapidly identify viable path types during initial stages. Simultaneously, the minimal parameter count of channel modulation significantly reduces computational overhead, providing an efficient lightweight solution for long-horizon audio-visual reinforcement learning training.
The conditional vector $c_t = [F_t^a; g_t]$ is input into the parameter generation network $\Psi$ to generate the corresponding scaling coefficients $\gamma$ and bias coefficients $\beta$ for each individual channel of the visual feature map $F_t^v$. Here, $\Psi$ is simply implemented as a standard multi-layer perceptron (MLP) to effectively learn these modulating parameters:
\begin{equation}
\gamma, \beta = \Psi(c_t).
\end{equation}
After spatial dimension broadcasting, these parameters perform a per-channel affine transformation on the visual features:
\begin{equation}
\tilde{F}_t^v = (1 + \gamma) \odot F_t^v + \beta.
\end{equation}
ACVF is a channel-based conditional mechanism rather than a spatial attention mechanism. $\gamma$ and $\beta$ are broadcast across spatial positions. This mechanism amplifies channels supporting the intention while suppressing less relevant channels, thereby generating a more directional fusion representation $\tilde{F}_t^v$ for the policy. This assists the agent in making more efficient decisions.

\section{Experiments}
\subsection{Experimental Setup}
We utilize Soundspaces \cite{b2} to evaluate our method, which comprises two independent environments: Matterport3D \cite{b22} and Replica \cite{b21}. To align with prior work \cite{b2}, we partition scenes into three groups: training/validation/test, consisting of 9/4/5 for Replica and 73/11/18 for Matterport3D. We evaluated our agent's performance across both environments under two distinct settings \cite{b2}: Heard and Unheard. The Unheard setting in Matterport3D is the most demanding and our primary focus. The strategy is trained from scratch. Our agent is optimized using Proximal Policy Optimization (PPO) across 5 parallel environments for 40{,}000 updates. We set the PPO hyperparameters as follows: clip parameter $0.1$, $4$ epochs per update, $1$ mini-batch, value loss coefficient $0.5$, entropy coefficient $0.20$, and maximum gradient norm $0.5$. The learning rate is $2.5\times10^{-4}$ with linear decay, and we use $\epsilon=1\times10^{-5}$.

\begin{table}[htbp]
\centering
\caption{Ablation Studies of the SDLD and ACVF Modules On the Unheard Setting.}
\label{tab:tab3}
\setlength{\tabcolsep}{8pt}
\renewcommand{\arraystretch}{1.2}

\begin{tabular}{lcccc}
\toprule
\multirow{2}{*}{\textbf{Model}} 
& \multicolumn{2}{c}{\textbf{Replica}} 
& \multicolumn{2}{c}{\textbf{Matterport3D}} \\
\cmidrule(lr){2-3}\cmidrule(lr){4-5}
& SR ($\uparrow$) & SPL ($\uparrow$)
& SR ($\uparrow$) & SPL ($\uparrow$) \\
\midrule
w/o SDLD and w/o ACVF 
& 50.9 & 34.7 & 33.5 & 21.9 \\
w/o SDLD 
& 66.1 & 38.6 & 56.2 & 37.8 \\
w/o ACVF 
& 67.6 & 36.7 & 56.2 & 39.5 \\
\textbf{SACF (Ours)} 
& \textbf{79.1} & \textbf{43.9} & \textbf{58.3} & \textbf{42.4} \\
\bottomrule
\end{tabular}

\vspace{2pt}
{\footnotesize \textit{Note: ``w/o'' denotes without.}}
\end{table}

We adopt standard SoundSpaces metrics \cite{b2} for evaluation: Success Rate (SR) measures the proportion of successful navigation; Success weighted by Path Length (SPL) evaluates path efficiency; and Success weighted by Normalized Alignment (SNA) assesses the directional alignment between agent actions and the sound source.

\begin{table}[htbp]
\centering
\caption{Parameter comparison of different audio-visual fusion mechanisms.}
\label{tab:tab4}
\setlength{\tabcolsep}{10pt}
\renewcommand{\arraystretch}{1.15}

\begin{tabular}{l|cc}
\hline
\textbf{Model}
& \textbf{Replica} 
& \textbf{Matterport3D} \\ \cline{2-3}
& Params (M)$\downarrow$
& Params (M)$\downarrow$ \\ \hline
Simple Concatenation  
& \textbf{4.35M} & 4.95M \\
Cross-Modal Spatial Attention 
& 7.06M & 7.67M \\
\textbf{SACF (Ours)} 
& 4.50M & \textbf{4.49M} \\
\hline
\end{tabular}
\end{table}

\subsection{Comparison Results}
We compare our approach with  baselines for audio-visual navigation. Table~\ref{tab:tab1} shows that our SACF framework, using depth maps as the visual input, achieves higher SPL and SR on the Replica dataset~\cite{b21} in both the Heard 80.3\%/96.3\% and Unheard 43.9\%/79.1\% splits. Compared to SoundSpaces, SACF improves SPL/SR by +5.9\%/+4.9\% on Heard and +9.2\%/+28.2\% on Unheard. On the SNA metric, SACF consistently outperforms SoundSpaces +3.6\% on Heard and +2.0\% on Unheard, but is slightly below AGSA on Replica 51.7\% vs.\ 52.0\% on Heard; 18.7\% vs.\ 22.4\% on Unheard. On Matterport3D~\cite{b22}, SACF yields improvements of +0.7\%, +1.3\%, and +1.2\% in SPL, SR, and SNA on Heard scenes, and larger gains of +20.5\%, +24.8\%, and +17.6\% on Unheard scenes. In particular, SACF attains 42.4\%/58.3\%/28.0\% in SPL,SR, and SNA on the Unheard split, indicating strong generalization to unseen sound sources. 
We attribute these gains to modeling sound source locations as discrete distributions over direction and distance, which provides compact spatial priors from auditory cues to guide the policy while suppressing irrelevant visual regions. The slightly lower SNA on Replica, compared to AGSA, suggests a trade-off between navigation efficiency and final bearing alignment.

\begin{figure}[htbp]
  \centering
  \includegraphics[width=\columnwidth]{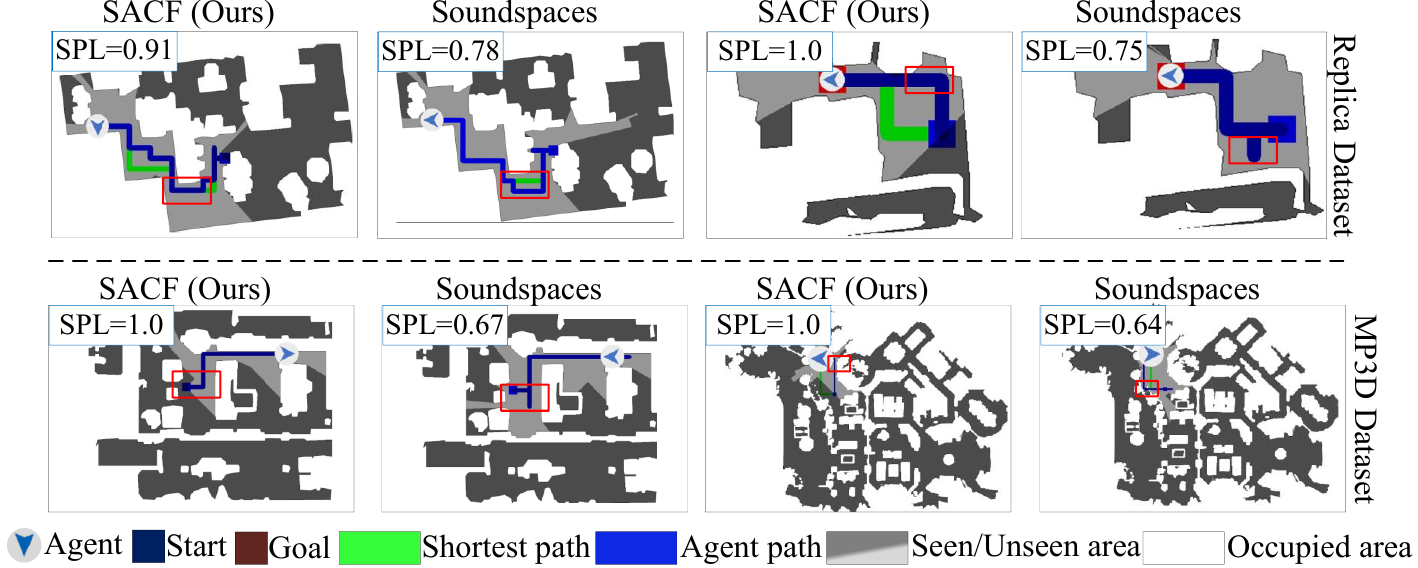}
  \caption{Red boxes highlight key turning points where acoustic cues successfully guide SACF to avoid dead ends and navigate more efficiently than the baseline.}
  \label{fig:fig4}
\end{figure}

As shown in Table~\ref{tab:tab2}, our SACF framework using RGB images as visual input also demonstrated improved navigation performance over the SoundSpaces[2] baseline. Since RGB images are susceptible to interference from lighting variations and texture details, depth maps directly provide geometric structural information about the environment. This enables agents to navigate obstacles more effectively and perceive spatial structures, making depth maps superior to RGB images for this task.

\subsection{Ablation experiment}
\paragraph{Module ablation}

To validate the contributions of each SACF component, we conduct ablation studies on Replica and Matterport3D datasets across three configurations, Baseline: Both SDLD and ACVF are removed; audio and visual features from standard encoders are simply concatenated. w/o SDLD: ACVF is retained, but the FiLM layer is conditioned only on global audio embeddings without spatial descriptor guidance. w/o ACVF: SDLD is retained, but its spatial descriptor $g_t$ is directly concatenated with unmodulated visual features, bypassing the channel modulation mechanism. As shown in Table~\ref{tab:tab3}, incorporating either module independently outperforms the baseline, confirming their individual effectiveness.

\begin{figure}[htbp]
  \centering
  \includegraphics[width=\columnwidth]{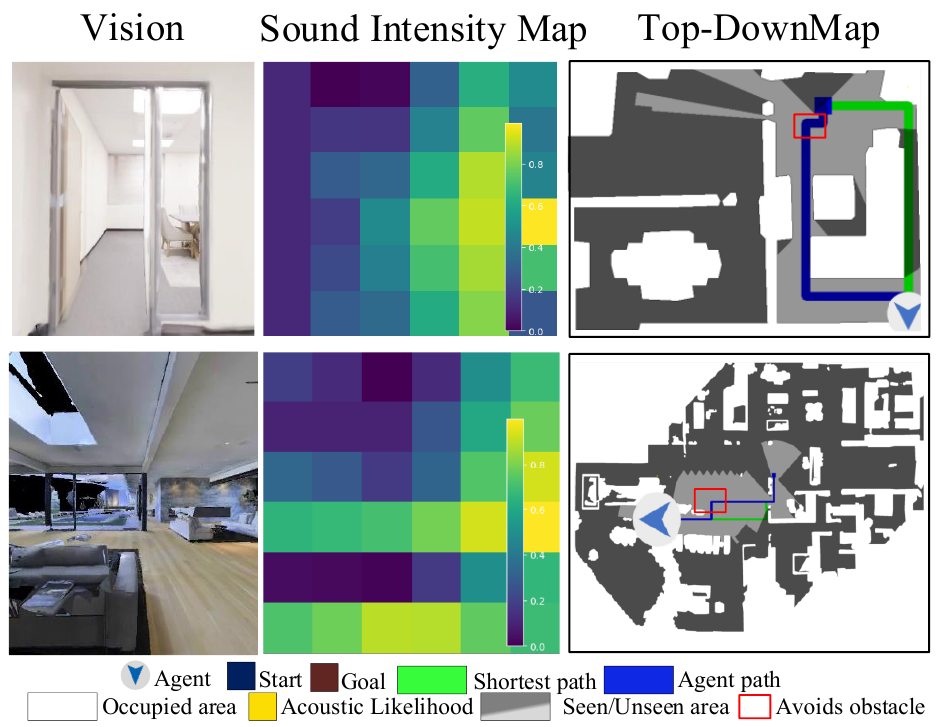}
  \caption{Top-down navigation trajectory diagram for navigation.}
  \label{fig:fig6}
\end{figure}

\paragraph{Parameter Comparison}

To demonstrate the lightweight advantages of ACVF, Table~\ref{tab:tab4} compares the efficiency of the three fusion schemes. Compared to Simple Concatenation, SACF adds only 0.15M parameters and is more compact than Cross-Modal Spatial Attention. Theoretically, unlike Cross-Modal Spatial Attention, which scales quadratically with spatial resolution, our FiLM-based ACVF operates with strictly linear computational complexity. Empirically, SACF maintains a high training throughput of approximately 48 FPS on Replica and approximately 74 FPS on Matterport3D. This ensures an extremely low real-time factor (RTF) during inference, confirming its lightweight nature and suitability for resource-constrained devices.

\subsection{Qualitative analysis}
\begin{figure}[htbp]
\includegraphics[width=1\columnwidth]{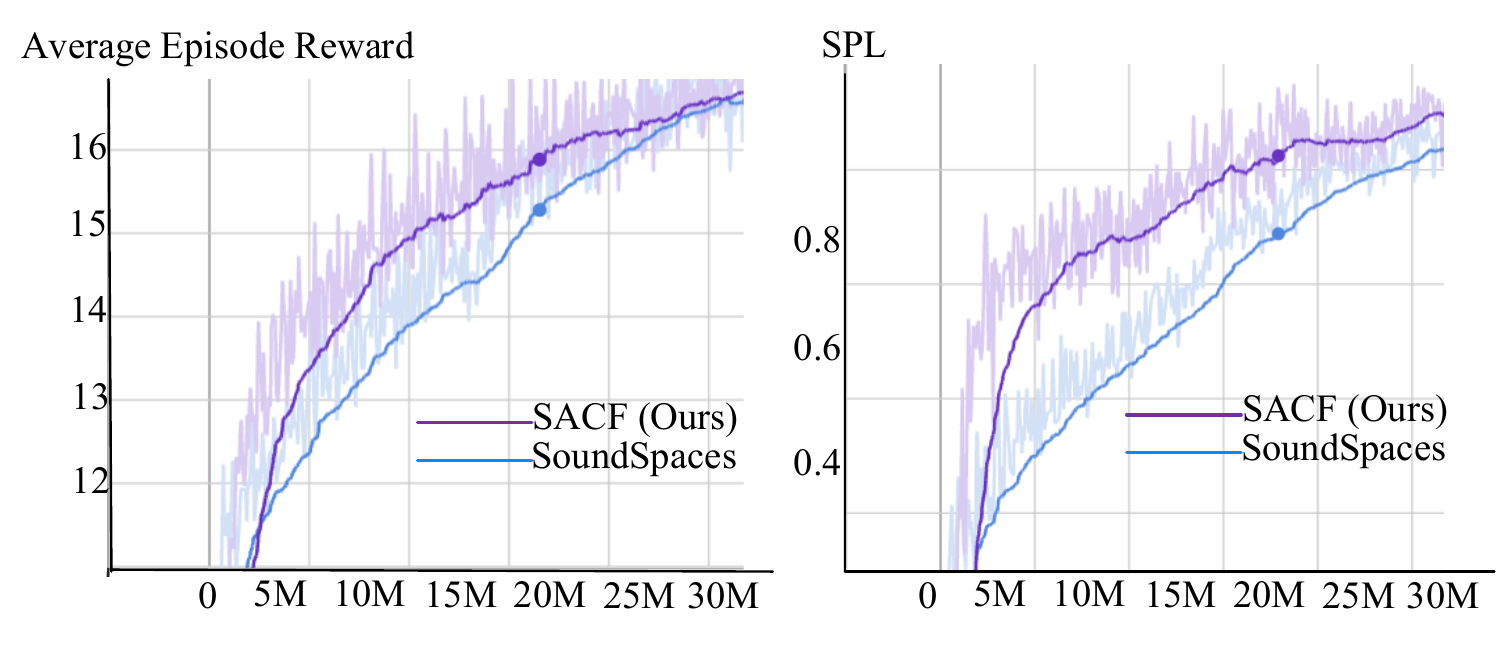} 
\caption{Convergence: rewards and SPL (0–1; tables in \%).}
\label{fig:fig5}
\end{figure}

To gain a more intuitive understanding of how our model operates, we conducted a qualitative analysis of several typical navigation trajectories. Fig.~\ref{fig:fig4} and~\ref{fig:fig6} illustrates the navigation trajectory on the top-down map of the SDLD model. The accompanying sound intensity map accurately pinpoints the direction of the sound source, effectively guiding the agent to navigate toward the target. SACF uses a single fixed random seed for training and evaluation, as long-term reinforcement learning in a 3D environment requires significant computational resources. However, the reward and SPL curves shown in ~\ref{fig:fig5}, converge smoothly and without oscillations, indicating that our proposed SACF framework exhibits stable learning dynamics. This also demonstrates that SACF achieves higher SPL values with fewer training steps, indicating that it converges faster than SoundSpaces.

\section{Conclusion}
This paper proposes the Spatial-Aware Conditioned Fusion (SACF) framework, which effectively addresses the core challenges in existing audio-visual navigation methods, namely the ambiguous spatial intent of targets and the lack of directionality in visual representations, through explicit spatial discretization and deep cross-modal feature modulation. Experimental results demonstrate that the strategy combining Spatially Discretized Localization Descriptor (SDLD) with Audio-Descriptor Conditioned Visual Fusion (ACVF) not only significantly improves navigation efficiency on Replica and Matterport3D datasets, but also exhibits strong cross-scene generalization capabilities, particularly when encountering unfamiliar target sounds, while achieving an optimal balance between computational overhead and navigation performance through its lightweight FiLM-based design. This approach establishes a new effective paradigm for constructing efficient and robust embodied audio-visual navigation systems.

\section*{ACKNOWLEDGMENT}

This research was financially supported by the National Natural Science Foundation of China (Grant No.: 62463029).

\bibliographystyle{IEEEtran}
\bibliography{references}

\end{document}